*Pre-service Teachers' Perception of using Mobile Devices in Teaching Climate Change in Primary Schools*

Dr.C.A. Obiefuna

carolobiefuna@yahoo.com

**Alvan Ikoku Federal College of Education, Owerri, Imo State**

**&**

**Prof. G.C.Offorma**

gchiofforma@gmail.com

**University of Nigeria, Nsukka, Enugu State.**

Abstract

The realities of climate change are gradually dawning on everyone including children. The need for a disaster reduction education requires the use of mobile technologies to identify some of the impact of climate change within an environment and create awareness on the dangers associated with climate change. Since the pre-service teachers will teach the primary school pupils, it is aptthat the use of mobile technologies should constitute part of their preparation while in training. This paper examined pre-service teachers' perception of using mobile technologies in teaching climate change in the primary school. One hundred and fifty (150) pre-service teachers in two Colleges of Education in the erosion disaster zones of Anambra and Imo States in the south eastern state of Nigeria were used for the study. Three research questions guided the study. The study utilized a survey approach to collect and analyze the data. The results from the study show that the pre-service teachers were confident that the use of mobile devices will createsignificant climate change awareness.However, the pre-service teachers saw the need for using mobile devices fin their preparation.Suggestions were made towards ensuring the integration of mobile technology literacy in the pre-service teacher education curriculum.

# *Introduction*

Climate change is becoming a topical issue worldwide. The aggressive and catastrophic nature in which the ecosystem is depleted creates some doubt, if enough energy and measures are in place to address the climate change menace. Some of the menace Euro News in Avwiri (2013) include severe drought, extreme hot/ cold weather conditions, flooding that have led to the destruction of lives and properties. In Nigeria for instance, flooding is experienced with greater velocity. Extraordinary calamities occur against each previous year. Thus, land, properties and human lives are lost annually to erosion menace. Offorma, (2013) for instance observed that the 2012 flooding in Nigeria led to the stagnation of economic and social activities. While, Odjugo (2010) noted that the evolving climate change coupled with increasing temperature has plunged some localities into experiencing extreme weather conditions. Egboka, Igbokwe in (Obiefuna, 2013) cited 49 communities in the South East Nigeria that were traumatized with about 2.300 gully erosions. The gullies are still active and have defied control measures from the Federal government, states and local communities.(Ezim.O, 2011).Discussing the impact of gully erosion, (Obiefuna, 2013)observed that children are more vulnerable to the climate change aftermath. In attestation to the above observation, UNICEF, (2012) for instance notes that the number of children affected by disasters range from 66.5 million per year in the late 1990s, to as many as 175 million.These observations are not only worrisome but also challengethe environmental experts on ways to ameliorate the situation. One strategy, the awareness education among school children has remained deficient. The school children are not considered as participants to the challenges of the environmental problems despite the advocacy on the need to develop strategic plans which emphasize the importance of recognizing children as key stakeholders in the policy making process and promoting environmental education. The nine year basic education curriculum in Nigeria for instance includedenvironmental education as one of the emerging issues that will lead to the attainment of Millennium Development Goals (MDGs) and National Economic Empowerment and Development Strategies. (NEEDS) The environmental education which will help toimprove the children's knowledge and commitment towards their environment supportsUNICEF's advocacy on educating a child on disaster risk empowermentof the child to use such knowledge.The UN convention on the rights of children in strong terms opine that children have right to participate in decisions that affects them. The climate change is one of such problems and requires the active participation of children. This paper examined the pre-service teachers' perception on the use of mobile devices in teaching climate change in the primary school.

**Climate change literacy advocacy**

The National policy on environment according to AGENCY (1998) made a provision on strategies that lead Nigeria into an era of social justice, self-reliance and sustainable development in the 21st Century. Some of the strategies include:

- review of curricula at all levels of the educational system and promote the formal study of environmental concepts and sciences;
- Boosting environmental awareness and education through the involvement of indigenous social structures, voluntary associations and occupational organizations.

The review of the curriculum from the revised edition of the UBE curriculum according(Danmole, 2011), in a similar way incorporated emergent curricular issues for which environmental education is one.

The climate change can be taught under environmental education. This willkeep the children informedabout their environment and make them active participants. The inclusion of environmental education will inculcate climate literacy in the children and encourage them in climate change intervention programme. Climate literacy according to Young involves one's ability to participate in a discourse about climate change and show full understanding of such terminologies as global warming, greenhouse gas emission and energy consumption. Climate change discourse and active participation will involveusing ICT to gain and disseminate information.PANT and HEEKS (nd).,Mobile devices such as the mobile phones can be used to gain knowledge and disseminate information on climate change.Climate change literacy equally involves the application of experiences using different school subjects. Young for instance was of the view that teaching climate change in a foreign language such as English language prevents students from comprehending that climate change is a local issue. Young's observation is right because in Nigerian school system learners read topics on erosion, HIV Aids for examination purposes rather than for lifelong education and problem solving. The non-attachment of value to emergent issues in the curriculum is because value, a standard that guides and determines action, attitudes towards objects and situations have not been properly inculcated in the learner. Yogesh (2008). Children need to learn about the impact of climate change and the posttraumatic stress that accompany it by exchanging ideas with affected children. This willmake the children have a feel of what a disaster area looks like and be in a position to appreciate the UN convention on their right to participate in decisions affecting them. Children's literacy on climate change will largely makethem attach value to emergency situations, especially when they learn how to use technological tools in identifying, monitoring and disseminating information on climate change. The use of mobile technologies is discussed in the next section

**Mobile Devices and Climate change**

The idea of boosting environmental awareness among children using mobile technologies is completely new in the average Nigerian classroom. The 21$^{st}$ century skills however require knowledge of technology.(Kukulska-Hulme (2006); Kukulska-Hulme, 2007) concept of mobile technologies include mobile devices such as cell phones,Personal media players, Personal Digital Assistants (PDAs), smart phones and wireless laptops.Winters, (2007) viewsthe following as the characteristics of mobile devices

- Enables knowledge building by learners in different context
- Enabling learners to construct understanding
- Enables change pattern of learning/work activities
- It is more than time and space

D. Laurillard (2007),As a mediating tool, the mobile technologies during the learning process address as follows:
- The learner and their personal relationship to the teacher and peer groups
- Relationship of the prior experience to the new topic and
- Where and when the learners are learning.

Winters and Laurrilard's observations on the mobile technologies call to mind some of the deficiencies encountered in use of the old technologies in teaching. There is nogainsaying that some of the old technologies were static and were time and space bound. They were more of teacher centered than learner centered. Knowledgeconstruction was hindered as a result of passive nature in which the teaching learning process situates the learner. Some of the old technologies may not have promoted personalized learning, authentic and collaborative learning whichKukulska-Hulme (2006) described as new forms of learning.

The use of the mobile devices among children is becoming a common phenomenon.Oksman and Rautiainen (2003) noted that young people have been particularly quick to adopt the mobile phones and the internet in their lives, and about 60% of children aged 9-12 in Finland own mobile devices. In a survey on the use of mobile devices for environmental education, Sulake in Ferry (2009) found out that teenagers use the mobile devices for taking pictures, sending text messages, listening to music and playing games. Bachmair (2007), notes that mobile devices such as PDAs and the mobile phones are used to access information from the web. Roberts (2008) in a digital shoebox project notes that children can use mobile phones to capture images from home, share them in an online environment and use the photographs as a resource materials.Willems (2012).citedinstances where mobile technologies, mobile apps and social networking were major sources of information during emergencies and after disaster.A major disadvantage in the use of mobile devices according to (MTEGA et al.) is the limiting screen size,and memory capacity, inability to use words, excel and power point functions.In Nigeria, the use of mobile phones among children is equally not new as most parents allow their children to make use of mobile devices. However, what may be new is the children's use of mobile devices in learning. Nigerian children can embark on projects on climate change using mobile technologies if they are properly guided. The observations on the use of mobile phones during emergency lay credibility that mobile phone can be allowed by school authority, rather than the reprimand against the use of cell phones The pre service teachers are to facilitate and guide the young learner in mobile phone usage, it may be righty assumed that by their training, the pre service teachers are competent in the use of mobile technologies in teaching the primary school children. The next section discusses the pre-service teacher competency in the use of mobile devices.

**Pre- Service Teachers and the Use of Mobile Devices**

The pre service teachers or the intern teachers are the future teachers who are expected to take over the leadership and management of primary school teaching on graduation. Theobjectives of the teacher education programme in section 8 article 71 of the National Policy on Education,(2004) include among others, providing motivated and conscientious teachers, who are creative, committed to the teaching profession and can adapt to the changing situations. One of such changing situations is competency in the use of Information and Communication Technology. The Federal government therefore in conformity with her promise on IT education as contained in the in the National Information Communication Technology Policy made ICT compulsory in schools, noted thus:

> In recognition of the prominent role of ICT in advancing knowledge and skills necessary for effective functioning in the modern world, there is need to integrate ICT into the Nigerian education system, and shall therefore provide basic infrastructures and training in realization of this goal. (NPE, 2004, p17).

The use of ICT by pre service teachers prepares them for the immediate and future challenges especially as Prensky, (2006) has described most 21 century children as digital natives who understand and speak in the language of technology.

ICT according toOlakulehin and Island (2007) facilitates interaction between and among individuals, provides more flexible and effective ways for lifelong professional development for today's teachers. Jung (2005) opines that ICT promotes active and autonomous learning,Smeets (2005) was of the view that ICT provides computer based tools for subject areas such as mathematics and technical drawing.The use of mobile technology is an aspect of ICT and requires the pre-service teachers' knowledge of how to store,retrieve and shareinformation. The pre-service teachers, in addition to their knowledge of ICT are exposed to general studies courses where they study emergent issues such as climate change. This preparation enables them meet the challenges such as the use ICT for climate change.

This paper examined the pre-service teachers' perception of using mobile technologies in teaching climate change.

### Statement of the Problem

The pre-service teachers are expected to teach primary school children. In the 21$^{st}$ century teaching and learning, they are expected to be literate in ICT.The high incidence on climate change devastation calls for creation of awareness through data collection, storing and retrieval of data, disseminating and sharing of information using mobile devices. The objective of this paper is to ascertain the pre-service teachers' perception of using mobile technologies to teach climate change in the primary school. Specifically, the objective is posed using the following research questions;

1. To what extent will the use of mobile devices in teaching climate change in the primary school create awareness?
2. To what extent will the use of mobile devices help in reducing climate change devastations?
3. To what extent are the pre-service teachers competent in the use of mobile devices?

**Methodology**

This is a descriptive study investigatng the pre-service teacher's perception of the use of mobile technologies in teaching climate change in the classroom. The area of the study is the south east geo political zone of Nigeria. The population of the study consists of all the pre-service teachers in the Colleges of Education in the South East of Nigeria.Two Colleges of Education were purposively sampled out of the five Colleges of Education. All the second year students in the selected Colleges of Education were used for the study,A total of one hundred and fifty students (150) were systematically sampled from three hundred and seventy five second year students qualified for the teaching practice exercise. A major instrument for data collection was the questionnaire.A twenty item semi structured questionnaire on Likert four point scale was administered to the students through the informed consent forms in keeping with. British Association Research Association (BERA) on ethical guidelines for educational research. The data were analyzed using the simple means. A mean score of 2.5 was accepted, while a mean score of below 2.5 was rejected. The results from the survey are discussed in the next section.

**Results**

Mean Responses

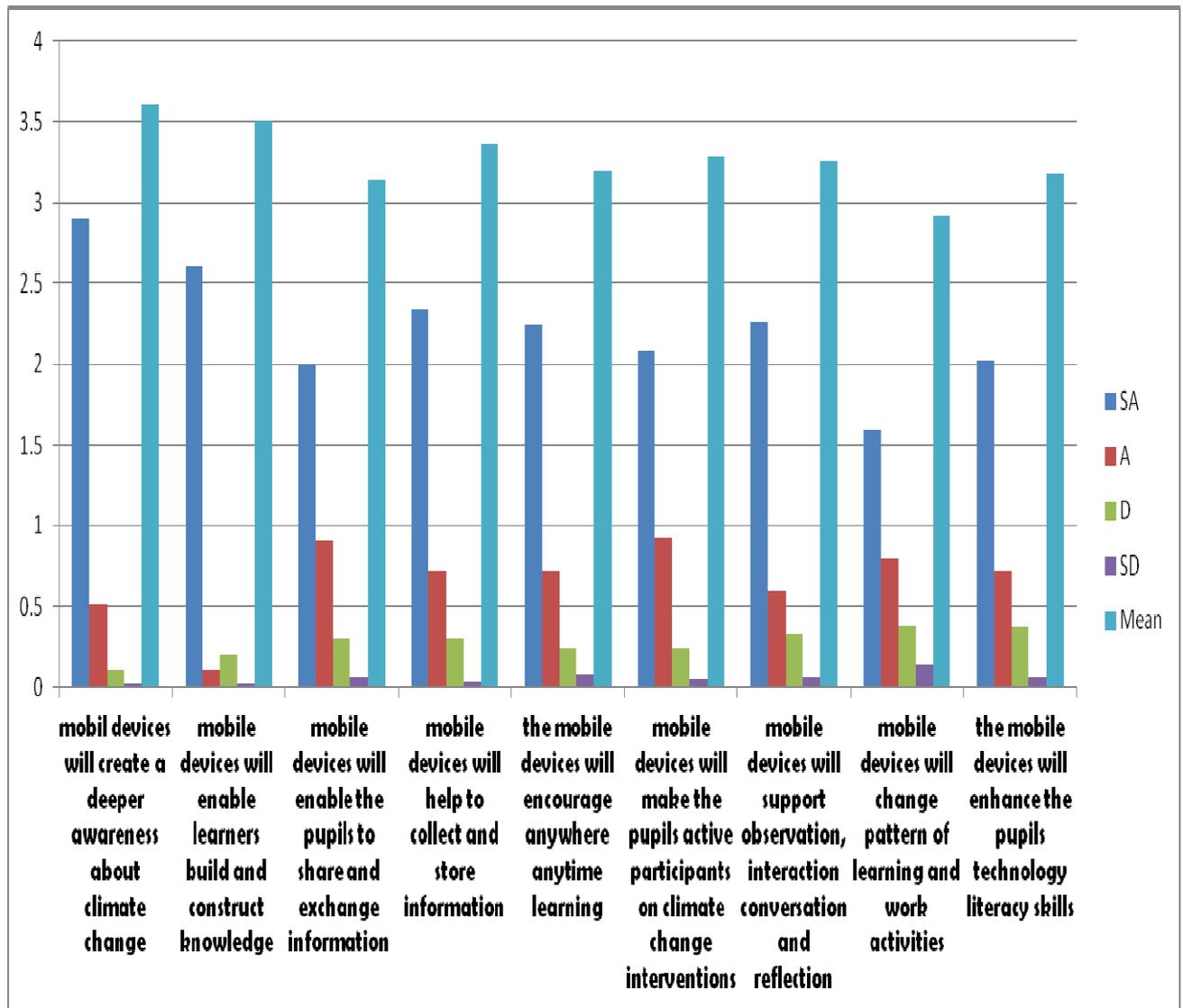

**Fig.1 . Responses of Pre-service Teachers' on using Mobile Devices in Teaching Climate change to create Awareness**

The different mean resposes in Fig. 1 above show that with 2.5 in each item indicate that using mobile devices to teach in the primary school can create climate change awareness.

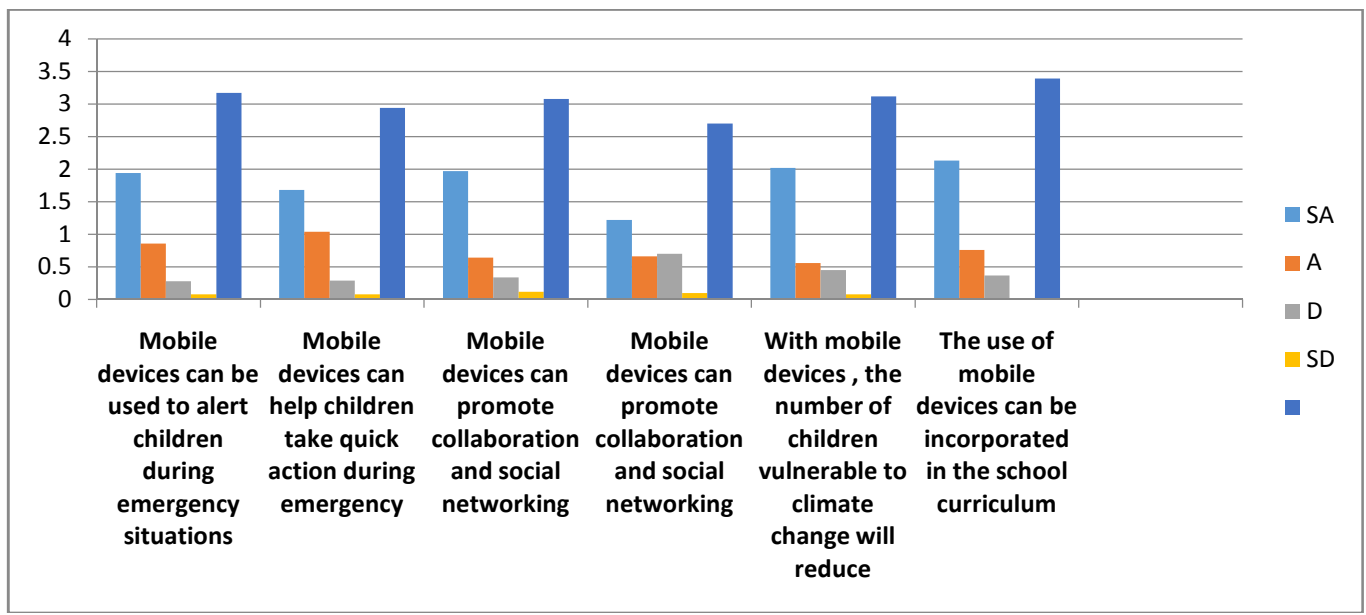

Fig 2: **Pre-service Teachers' Response on the use of mobile devices to reduce climate change devastations**

All the means in Fig2 are above 2.5 showing that mobile devices can be used in reducing climate change devastations.

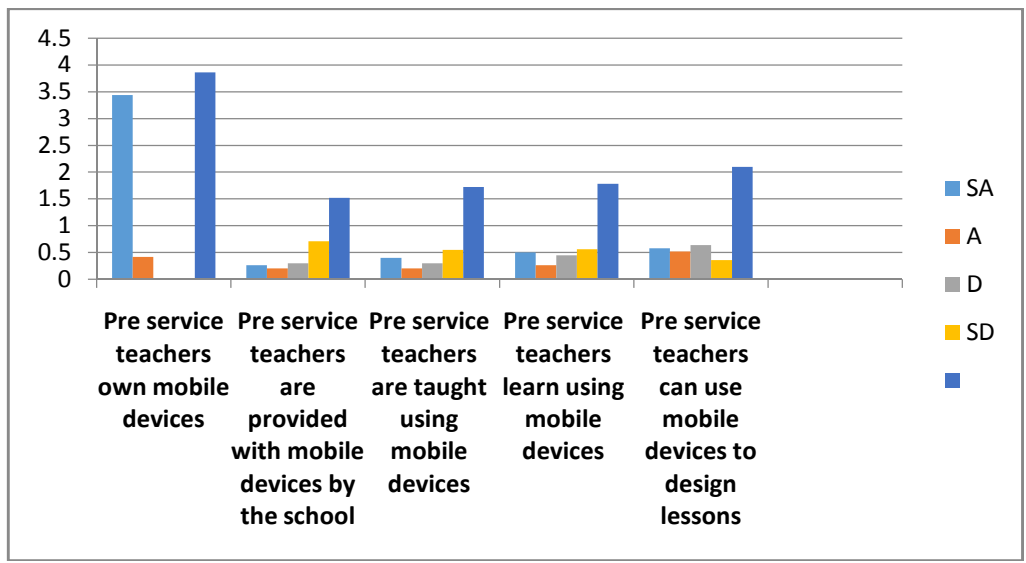

**Fig 3**. **Pre-service Teachers competency in the use of mobile devices in teaching climate change**

The result in Fig 3 shows majority of the means are below 2.5indicating that pre-servce teachers lack competence in the use of mobile devices in teaching climate change in the primary school.

**Summary of Findings**

The results from the study show that the pre-service teacher were of the opinion that the use of mobile devicesin teaching climate change in the primary schoolcreatessawareness and can help to prepare childrenduring emergency situations. thereby helping in the reduction of climate change devastation, The pre-service teachers,however, lack competence in the use of the mobile devices in teaching climate change in the primary school.

**Discussion**

Research Question1. Sought to find out the extent the use mobile devices in teaching climate change in the primary school can create awareness. In Fig1, the average means show that the pre-service teachers were of the view that the use of mobile devices in teaching climate change can create awareness. A mean score of 3.6 shows that the use of mobile devices can create a deeper awareness on climate change.Children can use mobile device such as the smart phones to collect pictures of erosion sites, flooded environments, gas flaring and upload such pictures on the internet and share with their peers in other environments. The children can also use text messages to inform and share ideas of the pictures with their peers. The uploaded pictures can trigger discussions and children can make contributions and exchange information on profering solutions..This interaction involves actual participation and hands on experience rather thanin the traditional classroom setting where children memorise lessons without actual participation, The discussion on the uploaded pictures supports Vygotsky's social learning theories which encourages the 4cs of creativity, connstruction,communication and collaborative learning,group work and discussion based learning. The discussion which emmanatesas a result of use of the mobile devices also supportWinters (2012) description of the charateristics of mobile technologyof knowledge consrtuction and building. The knowledge construction and building creats a deeper sense of value attachment unlike when the lesson is imagined and memorized.The discourse in knowledge building encourages active participation which according to UN convention on the rights of the children provide themwith opportunity to contribute to what affects them.

A mean score of 3.36 shows that with the use of mobile devices such as the smart phone or the camera, the children will be able to store and retrieve information and share such information with their peers. Information storage and retrivalhas been one of the challenges learners encounter. After lessons for instance,some learners hardly store or retrieve information from the long term memory and According to Thorne (2005) these group of children receive grades that do not match time and effort spent in studying. The use of mobile devices will improve information storage and retrieval in learners, especially with the mobility of the devices. thuslearning can take place both in the classroom and out of class, formally and informally.It supports and encouragesanywhere and anytime learning,change in classroom activity patterns and above all ability to personalise learning.(Diana Laurillard, 2007; Winters, 2012).

however, MTEGA et al.'s) observation on the limitations in memeory storage of the mobile devices is the major handicap, as limited information will be stored,or new ones replace the old.

Research Question 2. To what extent will the use of mobile technology help in reducing climate change devastations?

The finding in Fig 2 shows that with a mean of 3.05, the pre-service teachers' were of the opinion that the mobile devices will help to reduce climate change devastation. The use of the mobile devices for instance will help to alert children during emergencies and spur them into quick action such as alerting others, providing people outside the scenerio with information about their safety or help needs. Children's ability to upload pictures during crises will help to promote social networking.The pictures will serve as a first hand information of what happened and can lbe used forsharing ideas. The use of the mobile devices supports UNICEF' (2012) Willems (2012)views on emergency preparations towards climate change crisies or disaster. The pre-service teachers were of the view that the incorporation of mobile device as a mediating tool in the school curriculum will be applied in real life situationduring emergency. This is correct because the application will be based on problem solving advocated for in teaching and learning. Willem had earlier shown that the use of mobile devices reduced vunrability of earth quake victims as a result of transfer of applicaion of what was learnt in class into a real life situation.

Research Question 3. To what extent are the pre-service teachers competent in the use of mobile devices?

The result in Fig. 3. shows a grand mean of 2.19. This mean is below the expected mean and indicates incompetence of the pre-service teachers in the use of mobile devices. However,a mean of 3.86 show that a good number of the pre-service teachers own mobile devices. The pre-service teachers with means score of less than 2 show that the school does not provide them withmobile devices. They were not taught how to use the mobile devices neither do they learn using the mobile devics. The observation negates the policy satement of the Federal government on ensuring that ICT is incoporated in the school curriculum. If incorporation is based mainly on verbal statements without enforcing it by providing the necessary ICT materials and training of teacher educators who train the pre-service teachers, very little is expected from the pre-service teachers on graduation. Policy statements are backed up with facilities and infrastructures. There is need for the government to ensure the implementation of its policy statements.

The ownership of the mobile deivices by the pre-service teachers is a step forward into the effective utilization of the mobile devices. Students can be taught how to use mobile devices in teaching and learning to enable them prepare for the challenges they will meet in the labour market. Secondly, it will be absurd that primary school children are digitally informed but their teachers are not. The onus lies on the schools preparing the future teachers to ensure that the teaching and learning with ICT is made compulsory. This will enable the pre-service teachers appreciate the need for the use of ICT especially the mobile phones that isavailable and affordable.

## Conclusion

The study investigated pre-service teachers' perception on the use of mobille devices in teaching climate change in the primary school. Three research questions were used for the study. The results from the study show that the pre service teachers were optimistic that the use of the mobile devices in teaching climate change will to a great extent create awareness that will enable the children contribute and participate in the climate change reduction exercise. The use of the mobile phone will also enable the children take safety action especially when they are alerted during emergency situations.The pre-service teachers incompetence in the use of the mobile devices in teaching however is a major predicament. The preservice teachers were not taught how to use the mobile devices in teaching and learning and will find it difficult to teach the primary school children with mobile devices. Recommendations were made towards ensuring that the use of mobile devices in teachingwere incorporated in the pre-service teachers' training. The Federal government should also ensure that the promise made towards incorporationg ICT into the school curriculum isactualized as Nigeria strives to achieve vison 20:20 through Information and Communication Technology.

**EndNotes**

Appendix 1

| | Items | SA | A | D | SD | Mean | Dec. |
|---|---|---|---|---|---|---|---|
| | The use of mobile devices will create a deeper awareness about climate change | 448 | 78 | 16 | 4 | 3.6 | A |
| | The use of mobile devices will enable learners buildand construct knowledge | 392 | 99 | 32 | 3 | 3.5 | A |
| | The use of mobile devices will enable the pupils to share and exchange information on climate change | 296 | 126 | 40 | 10 | 3.14 | A |
| | The mobile devices will help to collect and store information | 352 | 102 | 46 | 5 | 3.36 | A |
| | The use of the mobile devices will encourage anywhere anytime learning | 336 | 108 | 36 | 12 | 3.2 | A |
| | The use of mobile devices will make the pupils active participants on climate change interventions | 312 | 138 | 36 | 8 | 3.29 | A |
| | The use of mobile devices will support observation, interaction conversation and reflection | 340 | 90 | 50 | 10 | 3.26 | A |
| | The use of mobile devices will change pattern of learning and work activities | 240 | 120 | 56 | 22 | 2.92 | A |
| | The use of the mobile devices will enhance the pupils technology literacy skills | 304 | 108 | 56 | 10 | 3.18 | A |
| | Grand Mean | | | | | 3.27 | |

Appendix 2

| ITEMS | SA | A | D | SD | Mean | Decision |
|---|---|---|---|---|---|---|
| Mobile devices can be used to alert children during emergency situations | 292 | 129 | 42 | 13 | 3.17 | A |
| Mobile devices can help children take quick action during emergency | 252 | 156 | 44 | 12 | 2.94 | A |
| Mobile devices can promote collaboration and social networking | 296 | 96 | 52 | 18 | 3.08 | A |
| Mobile devices can help children device remedial strategies | 184 | 99 | 112 | 15 | 2.7 | A |
| With mobile devices , the number of children vulnerable to climate change will reduce | 304 | 84 | 68 | 12 | 3.12 | A |
| The use of mobile devices can be incorporated in the school curriculum | 320 | 114 | 56 | 4 | 3.29 | A |
| Grand mean | | | | | 3.05 | |

APPENDIX 3

| S/N | ITEMS | SA | A | D | SD | Mean | Decision |
|---|---|---|---|---|---|---|---|
| 16 | Pre service teachers own mobile devices | 516 | 63 | 0 | 0 | 3.86 | A |
| 17 | Pre service teachers are provided with mobile devices by the school | 40 | 30 | 46 | 107 | 1.52 | R |
| 18 | Pre service teachers are taught using mobile devices | 60 | 36 | 80 | 83 | 1.72 | R |
| 19 | Pre service teachers learn using mobile devices | 76 | 39 | 68 | 84 | 1.78 | R |
| 20 | Pre service teachers can use mobile devices to design lessons | 88 | 78 | 96 | 54 | 2.1 | R |
| | Grand mean | | | | | 2.19 | |